\documentclass[prl, twocolumn,showpacs,preprintnumbers,amsmath,amssymb]{revtex4}

\usepackage{graphicx}
\usepackage{bm}

\usepackage{amsfonts}
\usepackage{amssymb}

\graphicspath{{.}{figures/}}

\newcommand\eqnref[1]{(\ref{#1})}
\newcommand\figref[1]{Fig.~\ref{#1}}

\newcommand{\iu}   {\mathrm{i}}     

\newcommand{\Omegarm}   {\mathrm{\Omega}}

\newcommand{\etal}   {\emph{et al}}
\newcommand{\ac} {\bot}   

\begin{document}

\title{Negative Refraction Requires Strong Inhomogeneity}


\author{Igor Tsukerman}%
\email{igor@uakron.edu}
\homepage{http://coel.ecgf.uakron.edu/~igor/}
\affiliation{%
Department of Electrical and Computer Engineering, The University of
Akron, OH 44325-3904
}%


\begin{abstract}
The paper establishes explicit lower bounds for the lattice cell
size of periodic structures (metamaterials and photonic crystals)
capable of supporting backward waves and producing negative
refraction. At optical frequencies, this result implies strong
inhomogeneity, in the sense that the cell size cannot be negligible
relative to the vacuum wavelength and the Bloch wavelength.
\end{abstract}

\pacs{41.20.-q, 41.20.Cv, 41.20.Jb, 42.25.Bs, 42.70.Qs}

\maketitle

%
%
Negative refraction (electromagnetic waves bending the `wrong' way
at material interfaces) and the closely related phenomenon of
backward waves (phase velocity at an obtuse angle with group
velocity) have become one of the most intriguing areas of research
in nanophotonics this century, with a number of books and review
papers readily available, e.g.
\cite{Milonni04,Eleftheriades05,Ramakrishna05,Pendry04,Shalaev06,Belov04},
and hundreds of research papers published. As early as in the 1940s,
Mandelshtam pointed out \cite{Mandelshtam47-50} that waves would
refract negatively at the interface boundary between a regular and a
backward-wave medium. In 1967, Veselago showed that media with
simultaneously negative (relative) dielectric permittivity
$\epsilon_r$ and magnetic permeability $\mu_r$ would support
backward waves and exhibit other unusual behavior of wave
propagation and refraction \cite{Veselago68}.

In 1999--2000, Pendry \emph{et al.} \cite{Pendry99} proved
theoretically and Smith \emph{et al.} \cite{Smith00} demonstrated
experimentally negative refraction in an artificial medium with
split-ring resonators. Furthermore, Pendry discovered that
Veselago's unusual `lens' -- a slab of a negatively refracting
material -- could produce a perfect image of a point source, thereby
beating the diffraction limit \cite{Pendry00}.

Truly homogeneous materials, in the Veselago sense, are not
currently known. Consequently, much effort has been devoted to the
development of artificial metamaterials capable of supporting
backward waves and producing negative refraction
\cite{Smith00,Shelby01,Parazzoli:PRL-03,Houck:PRL-03,Smith-SPIE04,
Shalaev06,Zhang:JOSA-06,Dolling:OptLett-06,Dolling:OptLett-07}.
Separately from the progress in metamaterials, negative refraction
has been observed and analyzed in singly and doubly periodic
waveguides \cite{Zengerle87} and in photonic crystals
\cite{Notomi00,Luo-PRB02,Foteinopoulou-PRB03,Cubukcu03,Parimi04,
Moussa05,Foteinopoulou-PRB05,Yannopapas05,Gajic05,Meisels06,Wheeler:PRB-06}.

All these intriguing findings have led to the presumption that there
are two species of negative refraction, one occurring in photonic
crystals and another one in metamaterials. Conceptually, the latter
are viewed as prototypical `Veselago media'.

There are, indeed, salient differences between metamaterials and
crystals in terms of the underlying structure, composition and
fabrication (e.g. lossless dielectric inclusions vs. lossy metallic
resonators of various kinds). On a more fundamental level, however,
all such structures can be characterized by a periodically varying
complex dielectric function, and from that point of view it is
legitimate to examine possible \emph{principal} differences between
metamaterials and photonic crystals. Importantly, can metamaterials,
as a matter of principle, be (arbitrarily) close to an ideal
homogeneous Veselago medium?

In the experimental and computational examples of the references
cited above, the cell size as a fraction of the vacuum wavelength
varies between $\sim0.11\div0.42$. One would hope that further
improvements in nanofabrication and design could bring the cell size
down to a smaller fraction of the wavelength, thereby approaching
the Veselago case of a homogeneous material. However, the main
conclusion of the present paper is that the cell size is constrained
not only by the fabrication technologies but by fundamental lower
bounds as well.


%
%
The analysis in this paper relies on the usual 2D and 3D renditions
of time-harmonic Maxwell's equations, and it is assumed that bulk
material parameters are applicable with a reasonable level of
accuracy. At optical frequencies, the relative intrinsic
permeability of all media can be set to unity (\cite{Landau84},
\S60; \footnote{Artificial magnetism can be created in periodic
dielectric structures at optical frequencies \cite{Cai07,Linden06}.
The equivalent `mesoscopic' permeability may then be different from
$\mu_0$, but the intrinsic \emph{microscopic} permeability of the
materials involved is still $\mu_0$.}.) To streamline the
mathematical development, we focus on square / cubic Bravais lattice
cells with size $a$ in 2D/3D and introduce dimensionless coordinates
$\tilde{x} = x/a$, etc., so that in these tilde-coordinates the 2D /
3D problem is set up in the unit square / cube. The $s$-mode in the
tilde-coordinates is described by the 2D wave equation
\begin{equation}\label{eqn:wave-eqn-Ez-normalized}
    \tilde{\nabla}^2 E ~+~ \tilde{\omega}^2 \epsilon_r E ~=~ 0,
\end{equation}
where $E$ is a one-component electric field phasor and
\begin{equation}\label{eqn:tilde-omega-defined}
    \tilde{\omega} = \frac{\omega a}{c}  ~=~ 2 \pi \, \frac{a}{\lambda_0}
\end{equation}
Here $c$ and $\lambda_0$ are the speed of light and the wavelength
in free space, respectively. The relative permittivity $\epsilon_r$
is a periodic function of coordinates over the lattice. In 3D, the
governing equation is
\begin{equation}\label{eqn:wave-eqn-vector-E-normalized}
    \tilde{\nabla} \times  \tilde{\nabla} \times \mathbf{E} ~=~
    \tilde{\omega}^2 \epsilon \mathbf{E}
\end{equation}
The fundamental solutions of the field equation in periodic
structures are known to be Bloch-Floquet waves with a (yet
undetermined) Bloch vector $\mathbf{K}_B$:
\begin{equation}\label{eqn:E-field-Bloch-wave}
    \mathbf{E}(\tilde{\mathbf{r}}) ~=~ \mathbf{E}_\mathrm{PER}(\tilde{\mathbf{r}}) \,
    \exp(\iu \tilde{\mathbf{K}}_B \cdot \tilde{\mathbf{r}})
\end{equation}
where $\tilde{\mathbf{r}}$ is the position vector. Subscript `PER'
implies periodicity with respect to any lattice vector $(n_x a, n_y
a, n_z a)$ with integer $n_x$, $n_y$, $n_z$ in the 3D case.
$\mathbf{E}_\mathrm{PER}$ can be expanded into a Fourier series
\begin{equation}\label{eqn:E-per-Fourier-expansion}
    \mathbf{E}_\mathrm{PER} (\tilde{\mathbf{r}}) ~=~
    \sum\nolimits_{\mathbf{n}}
    \tilde{\mathbf{e}}_{\mathbf{n}} \exp(\iu 2\pi \mathbf{n} \cdot \tilde{\mathbf{r}}),
\end{equation}
where $\tilde{\mathbf{e}}_{\mathbf{n}}$ are the Fourier coefficients
and index $\mathbf{n}$ runs over the integer lattice $\mathbb{Z}^2$
or $\mathbb{Z}^3$ in 2D/3D.

For analysis and physical interpretation of energy flow, phase
velocity and other properties of the Bloch wave, it is convenient to
view it as a suite of spatial Fourier harmonics (plane waves)
\cite{Lombardet05}. From \eqnref{eqn:E-field-Bloch-wave} and
\eqnref{eqn:E-per-Fourier-expansion},
\begin{equation}\label{eqn:E-eq-sum-harmonics-times-Bloch-factor}
    \mathbf{E}(\tilde{\mathbf{r}}) = \sum\nolimits_{\mathbf{n}}
    \mathbf{E}_\mathbf{n} \equiv \sum\nolimits_{\mathbf{n}}
    \tilde{\mathbf{e}}_\mathbf{n} \exp(\iu 2\pi \mathbf{n} \cdot \tilde{\mathbf{r}})
    \, \exp(\iu \tilde{\mathbf{K}}_B \cdot \tilde{\mathbf{r}})
\end{equation}
The decomposition of the magnetic field is similar.

%

%
%
It is important to note from the outset \cite{Lombardet05} that the
individual plane-wave components $\mathbf{E}_\mathbf{n}$ of the
electromagnetic Bloch wave do not satisfy Maxwell's equations in the
periodic medium and therefore do not represent physical fields. Only
taken together do these Fourier harmonics form a valid
electromagnetic field.

It is straightforward to verify that the plane waves in the
decomposition are orthogonal functions over the lattice cell (in the
sense of standard vector $\mathbf{L}_2$ inner product). Hence, by
Parseval's theorem, the time- and cell-averaged Poynting vector
$<\mathbf{P}> ~=~ \frac12<\mathrm{Re} \{ \mathbf{E} \times
\mathbf{H}^*\}>$ can be represented as the sum of the Poynting
vectors for the individual plane waves \cite{Lombardet05}:
\begin{equation}\label{eqn:Poynting-eq-sum-Pm}
    <\mathbf{P}> \,=\, \sum\nolimits_{\mathbf{n}} \mathbf{P}_\mathbf{n};
    ~~~~~  \mathbf{P}_\mathbf{n} \,=\, \frac{\pi n}{\tilde{\omega} \mu_0}
    \, |\tilde{\mathbf{e}}_\mathbf{n}|^2
\end{equation}
Group velocity $\partial \tilde{\omega} / \partial \tilde{k}$ is
clearly the same for all plane wave components, and hence group
velocity for the whole Bloch wave can be defined as $v_g = \partial
\tilde{\omega} / \partial \tilde{K}_B$.
%
%
In cases of weak dispersion, this velocity indeed approximately
represents signal velocity in the periodic medium
\cite{Yeh79,Tsukerman-book07}.

It is well known that in Fourier space the scalar wave equation
\eqnref{eqn:wave-eqn-Ez-normalized} becomes
\begin{equation}\label{eqn:Bloch-eigenvalue-prob-Fourier-domain}
    |\mathbf{K}_B + 2\pi \mathbf{n}|^2 \, \tilde{e}_\mathbf{n} ~=~
    \tilde{\omega}^2 \, \sum\nolimits_{\mathbf{m}}
    \tilde{\epsilon}_{\mathbf{n} - \mathbf{m}} \tilde{\mathbf{e}}_\mathbf{m}
\end{equation}
where $\tilde{\epsilon}_\mathbf{n}$  are the Fourier coefficients of
the dielectric permittivity $\epsilon$:
\begin{equation}\label{eqn:epsilon-Fourier-expansion}
    \epsilon ~=~ \sum\nolimits_{\mathbf{n}}
    \tilde{\epsilon}_\mathbf{n}
    \exp \left( \iu 2\pi \mathbf{n} \cdot \tilde{\mathbf{r}}
    \right)
\end{equation}
Indeed, the right hand side of
\eqnref{eqn:Bloch-eigenvalue-prob-Fourier-domain} is Fourier-space
convolution corresponding to real-space multiplication $\epsilon
\mathbf{E}$. The left hand side represents $-\tilde{\nabla}^2$.

Refraction at the interface between the periodic structure and air
(or another homogeneous dielectric) depends not only on the
\emph{intrinsic} characteristics of the Bloch wave in the bulk, but
also on the \emph{extrinsic} conditions at the interface boundary --
namely, the `excitation channel' \cite{Lombardet05}, i.e. the
Fourier component of the Bloch wave that couples to the incident
wave in the air \cite{Belov04,Lombardet05,Gajic05}. Intrinsic
properties include the forward or backward character of the wave --
that is, whether the Poynting vector and phase velocity (if the
latter can be properly defined) are at an acute or obtuse angle.

\begin{figure}
  \begin{center}
  \includegraphics[width=\linewidth]{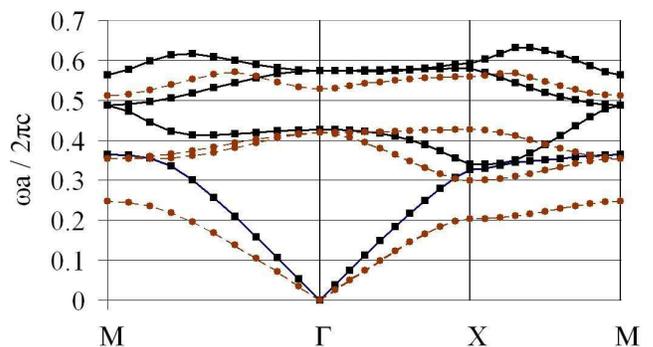}\\
  \caption{The photonic band diagram of the Gajic \etal~ crystal.
  TE modes ($p$-polarization, one-component $H$ field) --
  squares, solid lines. TM modes ($s$-polarization, one-component
  $E$ field) -- circles, dashed lines.}
  \label{fig:Gajic-PhC-band-diagram-TE-TM}
  \end{center}
\end{figure}

For illustration and further analysis, it is convenient to have a
specific example in mind (however, the analysis and conclusions will
be general). Consider the structure proposed by R.~Gajic, R.~Meisels
\etal~ \cite{Gajic05,Meisels06}. Their photonic crystal is a 2D
square lattice of alumina rods in air. The radius of the rod is
$r_\mathrm{rod}$ = 0.61~mm, the lattice constant $a$ = 1.86~mm, so
that $r_\mathrm{rod} /a \approx 0.33$. The band diagram, computed
using the plane wave method with 441 waves for $s$- and $p$-modes
appears in \figref{fig:Gajic-PhC-band-diagram-TE-TM} and, apart from
the scaling factors, is very close to the one in
\cite{Gajic05,Meisels06}, where various cases of wave propagation
and refraction are studied. In the context of this paper, of most
interest is negative refraction for small Bloch numbers in the
second band of the $p$-mode.

We observe that the TE2 dispersion curve is mildly convex around the
$\Gamma$ point ($K_B = 0$, $\omega a / 2\pi c \approx 0.427$),
indicating a negative group velocity for small positive $K_B$ and a
possible backward wave.

\begin{figure}
  \begin{center}
  \includegraphics[width=\linewidth]{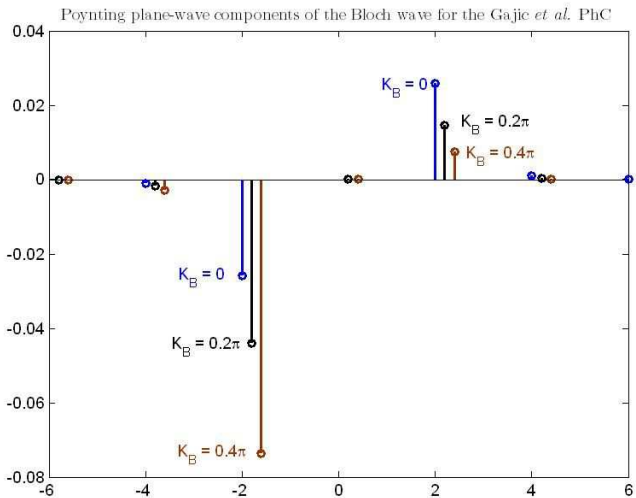}\\
  \caption{The plane-wave Poynting components $P_m$ for the Gajic
  \etal~ crystal (arb. units). Second $H$-mode (TE2) near the $\Gamma$ point
  on the $\Gamma \rightarrow X$ line.}
  \label{fig:Gajic-PhC-Poynting-harmonics-TE2}
  \end{center}
\end{figure}

An additional condition for a backward wave must also be satisfied:
the plane-wave component corresponding to the small positive Bloch
number must be appreciable (or better yet, dominant). The
distribution of Poynting components of the same wave is shown in
\figref{fig:Gajic-PhC-Poynting-harmonics-TE2}. It is clear from the
figure that the negative components outweigh the positive ones, so
power flows in the negative direction. (Details can be found in
\cite{Tsukerman-book07}.)

%

%
%
However, the normalized band diagram indicates that \emph{negative
refraction disappears in the homogenization limit} when the size of
the lattice cells tends to zero, provided that other physical
parameters, including frequency, are fixed. Indeed, the
homogenization limit is obtained by considering the small cell size
-- long wavelength limit $a \rightarrow 0$, $\tilde{K} \rightarrow
0$ (see \cite{Sjoberg05-149,Sjoberg05-760} for additional
mathematical details on Floquet-based homogenization theory for
Maxwell's equations). As these limits are taken, the problem and the
dispersion curves \emph{in the normalized coordinates} remain
unchanged, but the operating point $(\tilde{\omega},
\tilde{\mathbf{K}})$ approaches the origin along a fixed dispersion
curve -- the acoustic branch. In this case phase velocity in any
given direction $\hat{l}$, $\omega /K_l = \tilde{\omega} /
\tilde{K_l}$, is well defined and equal to group velocity $\partial
\omega / \partial K_l$ simply by definition of the derivative. No
backward waves can be supported in this regime.

This conclusion is not surprising from the physical perspective. As
the size of the lattice cell diminishes, the operating frequency
\emph{increases}, so that it is not the absolute frequency $\omega$
but the normalized quantity $\tilde{\omega}$ that remains
(approximately) constant. Indeed, a principal component of
metamaterials with negative refraction is a resonating element
\cite{Smith00,Smith-SPIE04,Ramakrishna05,Shalaev06} whose resonance
frequency is approximately inverse proportional to size
\cite{Linden06}.

It is pivotal in this paper to make a distinction between
\emph{strongly} and \emph{weakly} inhomogeneous cases of wave
propagation. The latter is intended to resemble an ideal `Veselago
medium,' with the Bloch wave being as close as possible to a
long-length plane wave. Toward this end, the following conditions
characterizing the weakly inhomogeneous backward-wave regime are put
forth:
\begin{itemize}
\item The first-Brillouin-zone component of the Bloch wave must
be dominant; this component then defines the phase velocity of the
Bloch wave.
\item
The other plane-wave components collectively produce energy flow at
an obtuse angle to phase velocity.
\item
The lattice cell size $a$ is small relative to the vacuum wavelength
$\lambda_0$;  $a / \lambda_0 \ll 1$.
\item At the air-material
interface, it is the long-wavelength, first-Brillouin-zone, plane
wave component that serves as the excitation channel for the Bloch
wave.
\end{itemize}

If any of the above conditions are violated, the regime will be
characterized as \emph{strongly inhomogeneous}: the EM wave can
``see'' the inhomogeneities of the material. By this definition, in
the weakly inhomogeneous case the normalized Bloch wavenumber
$\tilde{K}_B$ must be small, $\tilde{K}_B \equiv K_B a \ll \pi$.
Larger values of $K_B$ would indicate a strongly inhomogeneous (or,
synonymously, `photonic crystal' or `grating') regime, where the
lattice size is comparable with the Bloch wavelength. As we shall
see, under reasonable physical assumptions, backward waves cannot be
supported in the weakly inhomogeneous case; strong inhomogeneity is
required.

As a preliminary step in the analysis, it is instructive to examine
the direction of power flow for small $\tilde{K}_B$ in the lossless
cae (real $\epsilon$). The average Poynting vector is, according to
\eqnref{eqn:Poynting-eq-sum-Pm} and with a convenient normalization,
$$
    \tilde{P} ~\equiv~ 2 \tilde{\omega} \mu_0 <P> \,=\, K_B \,
    \left| \tilde{e}_0(\tilde{K}_B) \right|^2 ~+~
$$
\begin{equation}\label{eqn:Poynting-eq-sum-Pm-alt}
    \sum\nolimits_{m=1}^{\infty} (\tilde{K}_B + 2\pi m) \,
    \left| \tilde{e}_m(\tilde{K}_B) \right|^2
    ~+~ (\tilde{K}_B - 2\pi m) \, \left| \tilde{e}_{-m}(\tilde{K}_B) \right|^2
\end{equation}
The scalar form is used for notational convenience only; the
vectorial case is quite similar. It is, however, essential to
indicate explicitly that the Fourier amplitudes $\tilde{e}_m$ depend
on the Bloch parameter $\tilde{K}_B$. Since the waves corresponding
to $\pm \tilde{K}_B$ are complex conjugates of one another, we have
$\tilde{e}_{-m}(\tilde{K}_B) = \tilde{e}_{m}^*(-\tilde{K}_B)$, and
the expression for the Poynting vector becomes
$$
    \tilde{P} = \tilde{K}_B \left[ \left| \tilde{e}_0(\tilde{K}_B) \right|^2 +
    \sum\nolimits_{m=1}^{\infty} \left( \left| \tilde{e}_m(\tilde{K}_B) \right|^2
    + \left| \tilde{e}_{m}(-\tilde{K}_B) \right|^2 \right)
    \right]
$$
\begin{equation}\label{eqn:Poynting-eq-sum-Pm-pm-KB-temp}
   +~ 2\pi \sum\nolimits_{m=1}^{\infty} m \, \left( \left| \tilde{e}_m(\tilde{K}_B) \right|^2
   - \left| \tilde{e}_{m}(-\tilde{K}_B) \right|^2 \right)
\end{equation}
The first two terms in \eqnref{eqn:Poynting-eq-sum-Pm-pm-KB-temp}
are directly proportional to $\tilde{K}_B$. To make this small
parameter explicit in the third sum as well, we write
$$
   \tilde{P} \,=\, \tilde{K}_B \left[ \left| \tilde{e}_0 \right|^2 ~+~
    \, \sum\nolimits_{m=1}^{\infty} \left( \left| \tilde{e}_m(\tilde{K}_B) \right|^2
    + \left| \tilde{e}_{m}(-\tilde{K}_B)  \right|^2 \right) \right.
$$
\begin{equation}\label{eqn:Poynting-eq-sum-Pm-pm-KB}
   \left.
   +~ 2\pi \sum\nolimits_{m=1}^{\infty} m \,\,
   \frac{\partial \left| \tilde{e}_m \right|^2}{\partial \tilde{K}_B}
   \right]
\end{equation}
For small $\tilde{\omega}$, the positive term $|\tilde{e}_0|^2$ in
the square brackets tends to be dominant, making it difficult to
produce a negative power flow and a backward wave. This is so
because the magnitudes of all spatial harmonics \emph{except for
$\tilde{e}_0$} are for small $\tilde{\omega}$ constrained by
\eqnref{eqn:Bloch-eigenvalue-prob-Fourier-domain}:
%
%
\begin{equation}\label{eqn:em-via-tilde-omega}
    |\tilde{e}_n| ~\leq~ \tilde{\omega}^2
    \left| \tilde{\mathbf{K}}_B + 2 \pi \mathbf{n} \right|^{-2}
    \left\| \tilde{\epsilon} \right\|_{l_2}, ~~~ \left\| \tilde{e} \right\|_{l_2} = 1, ~~ n \neq 0
\end{equation}

The arguments above suggest that there must be a lower bound for the
relative cell size $a / \lambda_0 = \tilde{\omega} / 2\pi$ when the
medium could still support backward waves. To the best of my
knowledge, this question has not so far been posed explicitly in the
literature.

In the remainder, we investigate the constraints on the periodic
\emph{in the weakly inhomogeneous backward-wave regime}. This
implies that $\tilde{K}_B = K_B a \ll 1$. To simplify mathematical
analysis, we focus on the limiting case $K_B = 0$, but the
conclusions will apply, by physical continuity, to small
$\tilde{K}_B$. We first turn to the $s$-mode governed by the 2D
equation \eqnref{eqn:wave-eqn-Ez-normalized}.
%
%
For $\tilde{\omega} \neq 0$ and $\eta = \tilde{\omega}^{-2}$,
\begin{equation}\label{eqn:epsilon-E-eq-eta-nabla2-E}
    \epsilon E ~=~ -\eta \tilde{\nabla}^2 E
\end{equation}
%
Further analysis relies on the inversion of $\tilde{\nabla}^2$. To
do this unambiguously, let us split $E$ up into the zero-mean term
$E_{\ac}$ and the remaining constant $E_0$:
$
    E ~=~ E_0 + E_{\ac}
$.
Symbol `$\bot$' indicates orthogonality to the null space of the
Laplacian (i.e. to constants). To eliminate the constant component
$E_0$, we integrate \eqnref{eqn:epsilon-E-eq-eta-nabla2-E} over the
lattice cell. Integrating by parts and noting that the boundary term
vanishes due to the periodic boundary conditions ($K_B = 0$), we get
$$
    E_0 ~=~ -{\tilde{\epsilon}_0}^{-1} \,
    \int\nolimits_{\Omegarm} \epsilon E_{\ac} \, d \Omegarm,
    ~~~~~~ \tilde{\epsilon}_0 \neq 0
$$
(The exceptional case $\tilde{\epsilon}_0 = 0$ is mathematically
quite intricate and may consitute a special topic for future
research.) With $E_0$ eliminated, the eigenvalue problem for
$E_{\ac}$ becomes
$$
    \epsilon \, \left[ E_{\ac} \,-\,
    {\tilde{\epsilon}_0}^{-1} \, \int_{\Omegarm} \epsilon \, E_{\ac}
    \, d \Omegarm \right] ~=~ -\eta \tilde{\nabla}^2 E_{\ac}
$$
Since $E_{\ac}$ by definition is zero-mean,
\begin{equation}\label{eqn:inv-nabla2-eps-Eac}
    \tilde{\nabla}^{-2}_{\bot} \, \left\{ \epsilon \, \left[ E_{\ac} \,-\,
    {\tilde{\epsilon}_0}^{-1} \, (\epsilon, \, E_{\ac})
    \right] \right\} ~=~ -\eta  E_{\ac}
\end{equation}
where $\tilde{\nabla}^{-2}_{\bot}$ is the zero-mean inverse of the
Laplacian. Fourier analysis easily shows that this inverse is
bounded (the Poincar\'{e} inequality):
$
    \left\| \tilde{\nabla}^{-2}_{\bot} \right\| ~\leq~ (4\pi^2)^{-1}
$
Then, taking the norm of both sides of
\eqnref{eqn:inv-nabla2-eps-Eac}, we get
\begin{equation}\label{eqn:eta-leq-epsmax-eps0}
    \left| \eta \right| ~\leq~ (4\pi^2)^{-1} \, |\epsilon|_{\max}  \left( 1 \,+\,
    |\epsilon|_{\max} / \left| \tilde{\epsilon}_0  \right| \right)
\end{equation}
This result, that can be viewed as a generalization of the
Poincar\'{e} inequality to cases with variable $\epsilon_r$, leads
to a simple lower bound for the lattice cell size, with the mean and
maximum values of $\epsilon$ as parameters:
\begin{equation}\label{eqn:a-over-lambda0-geq-inv-eps-max}
    \left( \frac{a}{\lambda_0} \right)^2 ~=~ \frac{\tilde{\omega}^2}{4\pi^2}
    ~\geq~ \frac{1}{|\epsilon|_{\max}  \left( 1 \,+\,
    |\epsilon|_{\max} / \left| \tilde{\epsilon}_0  \right| \right)}
\end{equation}
%
%
%

Turning now to the vector field formulation
\eqnref{eqn:wave-eqn-vector-E-normalized}, we deal with 2D and 3D
cases simultaneously and rewrite the field equation as
\begin{equation}\label{eqn:epsilon-E-eq-eta-curl-curl-E}
    \epsilon \, \mathbf{E} ~=\, -\eta \,
    \tilde{\nabla} \times \tilde{\nabla} \times \mathbf{E}
\end{equation}
The $\tilde{\nabla} \times \tilde{\nabla} \times $ operator can be
inverted unambiguously if the result, denoted with $(\tilde{\nabla}
\times)^{-2}_{\bot}$, is sought in the functional space
$\mathbf{H}^{1}_{\ac}(\Omegarm)$ of divergence-free zero-mean
fields. For any such field $\mathbf{u}$,
%
$
    \tilde{\nabla} \times \tilde{\nabla} \times \mathbf{u} \,=\,
    -\tilde{\nabla}^2 \mathbf{u}
$
and hence
$$
    \left\| \tilde{\nabla}^2 \mathbf{u} \right\|_2^2 ~=~
    (\tilde{\nabla} \times \tilde{\nabla} \times \mathbf{u}, \,
    \tilde{\nabla} \times \tilde{\nabla} \times \mathbf{u} )
$$
$$
    =~  (\tilde{\nabla}^2 \mathbf{u}, \, \tilde{\nabla}^2 \mathbf{u})
    ~\geq~ (4\pi^2)^2 \, (\mathbf{u}, \, \mathbf{u})
    ~=~  (4\pi^2)^2 \,  \left\| \mathbf{u} \right\|_2^2
$$
This implies that the inverse curl-curl, considered as an operator
with its range in $\mathbf{H}^{1}_{\ac}$, is bounded:
\begin{equation}\label{eqn:inv-curl-curl-leq-4pi2-1}
    \left\| (\tilde{\nabla} \times)^{-2}_{\bot} \right\|  ~\leq~ (4\pi^2)^{-1}
\end{equation}
%
The relevant splitting of $\mathbf{E}$ is into the zero-mean
divergence-free term $\mathbf{E}_{\ac} \in \mathbf{H}^{1}_{\ac}$ and
the curl-free remainder \footnote{Curl-free fields are representable
as gradients if the domain is simply connected; this is certainly
true for any Bravais lattice cell.} $\mathbf{E}_0 = -\nabla \phi_0$
(the Helmholtz decomposition):
$
    \mathbf{E} ~=~ \mathbf{E}_{\ac} \,-\, \tilde{\nabla} \phi_0
$.
Field $\mathbf{E}_{\ac}$ is in fact, up to the factor $\iu
\tilde{\omega}$, the magnetic vector potential with the Coulomb
(zero-divergence) gauge.

Taking divergence (in the distributional sense) of the governing
equation \eqnref{eqn:epsilon-E-eq-eta-curl-curl-E} and integrating
over the cell, one eliminates the electrostatic term $\tilde{\nabla}
\phi_0$ and arrives at an eigenvalue problem for $\mathbf{E}_{\ac}$:
%
%
%
%
%
$$
    \epsilon \left( \mathbf{E}_{\ac} \,-\,
    \nabla \mathcal{L}_{\epsilon}^{-1} \,
    \tilde{\nabla} \cdot (\epsilon \mathbf{E}_{\ac})
    \right) ~=\, -\eta \tilde{\nabla} \times \tilde{\nabla} \times \mathbf{E}_{\ac}
$$
assuming that the electrostatic operator $\mathcal{L}_{\epsilon} =
\tilde{\nabla} \cdot \epsilon \tilde{\nabla}$ is nonsingular.
Equivalently, since $\mathbf{E}_{\ac}$ is by definition
divergence-free and zero-mean,
\begin{equation}\label{eqn:nabla-times-inv-epsE-leq-eta-E}
    (\tilde{\nabla} \times)^{-2}_{\bot} \,
    \left\{ \epsilon \left[ \mathbf{E}_{\ac} \,-\,
    \nabla \mathcal{L}_{\epsilon}^{-1} \,
    \nabla \cdot (\epsilon \mathbf{E}_{\ac})
    \right] \right\} ~=~ -\eta \mathbf{E}_{\ac}
\end{equation}
%
%
an upper bound for $\eta$ can be obtained by taking the
$\mathbf{L}_2$-norms of both sides, with
\eqnref{eqn:inv-curl-curl-leq-4pi2-1} in mind:
%
\begin{equation}\label{eqn:eta-leq-4pi2-epsmax}
    | \eta | ~\leq~ (4\pi^2)^{-1} |\epsilon|_{\max} \left(1
    \,+\, |\lambda|_{\max} (\mathcal{L}_{\epsilon}^{-1})
    |\epsilon|_{\max} \right)
\end{equation}
This estimate is analogous to the scalar one
\eqnref{eqn:eta-leq-epsmax-eps0}, except that the maximum eigenvalue
$|\lambda|_{\max} (\mathcal{L}_{\epsilon}^{-1})$ (not to be confused
with the wavelength) appears instead of the inverse mean value
$|\tilde{\epsilon}_0|^{-1}$ of the permittivity. This eigenvalue is
bounded unless the operating frequency is close to the quasi-static
plasmon resonance value. In the most general situation, no simple
estimate of $|\lambda|_{\max} (\mathcal{L}_{\epsilon}^{-1})$ is
available, but it can be computed numerically using a number of
algorithms (see e.g. \cite{Tsukerman-book07}) for any given
distribution of $\epsilon$ in the lattice cell.

At the same time, there are practically important situations where
the bound for $\eta$ can be made more explicit. One such case is
that of non-plasmonic materials, when $\epsilon \geq \epsilon_{\min}
> 0$ throughout the lattice cell. Then
$$
    |\lambda|_{\max} (\mathcal{L}_{\epsilon}^{-1}) ~\leq~
    \epsilon_{\min}^{-1} \, |\lambda|_{\max} (\nabla^{-2}_{\bot})
    ~\leq~ (4\pi^2 \epsilon_{\min})^{-1}
$$
and from the estimate \eqnref{eqn:eta-leq-4pi2-epsmax} for $|\eta|$
the following bound on the normalized cell size emerges:
%
%
\begin{equation}\label{eqn:a-over-lambda0-geq-epsmax-epsmin}
    \left( \frac{a}{\lambda_0} \right)^2 =
    \left( \frac{\tilde{\omega}}{2\pi} \right)^2 =
    \frac{1}{4\pi^2 | \eta |}  \geq
    \frac{1} { |\epsilon|_{\max} \left(1
    + |\epsilon|_{\max} / (4\pi^2 \epsilon_{\min}) \right)
    }
\end{equation}

\begin{figure}
  \begin{center}
  \includegraphics[width=\linewidth]{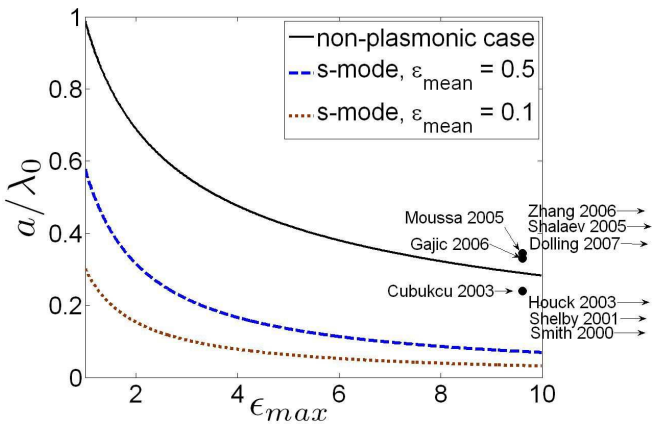}\\
  \caption{Bounds on the normalized cell size and a few representative
  data points from the literature.}
  \label{fig:cell-size-bounds}
  \end{center}
\end{figure}

The lower bounds \eqnref{eqn:a-over-lambda0-geq-inv-eps-max} and
\eqnref{eqn:a-over-lambda0-geq-epsmax-epsmin} are plotted as a
function of $|\epsilon|_{\max}$ in \figref{fig:cell-size-bounds},
for $\epsilon_{\min} = 1$ and two values of $\tilde{\epsilon}_0$
(0.5 and 0.1). For illustration, several representative data points
(both theoretical and experimental) from the literature are also
shown in the figure. In the microwave regime, when metals are very
good conductors and consequently $|\epsilon|_{\max}$ is high, the
theoretical bound for the cell size is non-restrictive and the
respective data points (Smith, Shelby, Houck, and others) easily
turn up above the relevant theoretical curve; these points lie off
the chart in \figref{fig:cell-size-bounds}.

The Cubukcu \etal~ data point lies \emph{below} the theoretical line
\eqnref{eqn:a-over-lambda0-geq-epsmax-epsmin}; however, there is no
contradiction because in this instance negative refraction occurs in
the vicinity of the M point, where the Bloch wavelength and the
lattice cell size are comparable. This constitutes, by our
definition, a strongly inhomogeneous case to which the theoretical
bound does not apply.

The Moussa and Gajic data points for non-metallic crystals lie only
slightly above the theoretical bound, indicating that this bound can
be approached in some cases. Still, it must be stressed that the
theoretical limits on the cell size are \emph{necessary}, but in
general \emph{not sufficient}, conditions for negative refraction. A
sufficiently large lattice cell size makes it \emph{possible} for
higher-order Fourier harmonics of the Bloch wave to outweigh the
first-Brilloin-zone harmonic, but does not guarantee that they will
do so and that they will have the desirable sign.

Another case where the theoretical bound
\eqnref{eqn:eta-leq-4pi2-epsmax} can be made more explicit is that
of a lossless host medium, $\epsilon = \epsilon_h$, with embedded
`inclusions' $\epsilon = \epsilon_i = \epsilon'_i + \iu
\epsilon''_i$ (spheres, split-ring resonators, fishnets, horseshoes,
rods, etc.). The eigenvalue $|\lambda|_{\max}
(\mathcal{L}_{\epsilon})^{-1} = |\lambda|_{\min}^{-1}
(\mathcal{L}_{\epsilon})$ can be estimated from the electrostatic
energy functional
$$
    (\epsilon \nabla \phi, \, \nabla \phi) ~=~
    \epsilon_h W_h \,+\, (\epsilon'_i + \iu \epsilon''_i) W_i
$$
where $
    W_{h,i} = \int_{\Omegarm_{h,i}} |\nabla \phi|^2 \, d \Omegarm
$.
%
%
Then $|(\epsilon \nabla \phi, \, \nabla \phi)|^2$ is a quadratic
form with respect to $W_{h,i}$ and can be bounded by direct
evaluation of its minimum eigenvalue. The end result, for small
losses $\epsilon''_i \ll \epsilon_h + |\epsilon_i|$, is
%
%
%
$$
    |\lambda|_{\min}^2 (\mathcal{L}_{\epsilon}) ~\gtrsim~
    \frac{ \epsilon_h^2 {\epsilon''_i}^2}
    {2 (\epsilon_h^2 + |\epsilon_i|^2)}
$$
This estimate can be used in conjunction with the general bound
\eqnref{eqn:eta-leq-4pi2-epsmax}.

In summary, it has been proved that periodic structures capable of
supporting backward waves and producing negative refraction in the
optical range must be strongly inhomogeneous. More precisely, the
lattice cell size, as a fraction of the vacuum wavelength and/or the
Bloch wavelength, must be above certain thresholds established in
this paper. These thresholds contain the maximum, minimum and mean
values of the complex dielectric permittivity as key parameters. In
the presence of good conductors (e.g. at microwave frequencies) such
theoretical constraints are not very restrictive. However, at
optical frequencies and/or for non-metallic structures the bounds on
the cell size must be honored and may help to design metamaterials
and photonic crystals with desired optical properties.

\bibliographystyle{plain}


\end{document}